\documentclass[numberedappendix]{emulateapj}
\begin{document}

\slugcomment{Accepted for publication in ApJ: March 20, 2008}

\title{A Resolved Molecular Gas Disk around the Nearby A Star 49 Ceti}

\author{A. M. Hughes \altaffilmark{1}, 
D. J. Wilner \altaffilmark{1}, 
I. Kamp \altaffilmark{2}, 
M. R. Hogerheijde \altaffilmark{3}}

\email{mhughes@cfa.harvard.edu}

\altaffiltext{1}{Harvard-Smithsonian Center for Astrophysics,
  60 Garden Street, Cambridge, MA 02138}
\altaffiltext{2}{Kapteyn Astronomical Institute, University of Groningen, 
  9700 AV Groningen, The Netherlands}
\altaffiltext{3}{Leiden Observatory, Leiden University, P.O. Box 9513, 2300 RA,
  Leiden, The Netherlands}


\bibliographystyle{apj}

\begin{abstract}

The A star 49 Ceti, at a distance of 61 pc, is unusual in retaining a 
substantial quantity of molecular gas while exhibiting dust properties similar 
to those of a debris disk.  We present resolved observations of the disk 
around 49 Ceti from the Submillimeter Array in the J=2-1 rotational transition 
of CO with a resolution of 1.0$\times$1.2 arcsec.  The observed emission 
reveals an extended rotating structure viewed approximately edge-on and clear 
of detectable CO emission out to a distance of $\sim 90$~AU from the star.  No 
1.3 millimeter continuum emission is detected at a $3\sigma$ sensitivity of 
2.1 mJy/beam.  Models of disk structure and chemistry indicate that the inner 
disk is devoid of molecular gas, while the outer gas disk between 40 and 200 AU 
from the star is dominated by photochemistry from stellar and interstellar 
radiation.  We determine parameters for a model that reproduces the basic 
features of the spatially resolved CO J=2-1 emission, the spectral energy 
distribution, and the unresolved CO J=3-2 spectrum. We investigate 
variations in disk chemistry and observable properties for a range of 
structural parameters. 49 Ceti appears to be a rare example of a system in a 
late stage of transition between a gas-rich protoplanetary disk and a tenuous, 
virtually gas-free debris disk.


\end{abstract}

\section{Introduction}

A key to understanding the formation of planetary systems lies in 
characterizing the transitional phase between the gas-rich primordial 
disks found around young T Tauri stars and the tenuous, virtually 
gas-free debris disks around their main-sequence counterparts.  Unfortunately, 
disks in this transitional phase are rare and difficult to identify.
Dust disks around young stars are commonly identified through the 
``Vega-excess'' phenomenon (first observed using the Infrared 
Astronomical Satellite by \citealt{aum84}; see review by \citealt{zuc01}), 
in which an infared excess over the stellar photosphere is attributed to 
reprocessing of optical and ultraviolet 
starlight by thermally emitting circumstellar dust grains.  
49 Ceti was first identified in this way by \citet{sad86}.
The quantity $\tau = L_{IR}/L_{bol}$  is often used as an indicator of the 
``optical depth'' of the dust disk, as it provides a rough estimate of the 
quantity of optical/ultraviolet light intercepted and reemitted by the dust.  
\citet{jur93} correlated the IRAS Point Source Catalog with the Yale Bright 
Star Catalog \citep{hof91} and identified three A stars with 
$\tau > 10^{-3}$, indicative of tenuous, optically thin circumstellar dust.  
Two were the stars $\beta$ Pic and HR4796, which are now known to host 
debris disks.  The third was 49 Ceti, which unlike the other two defies 
classification as a debris disk because it retains a substantial quantity of 
molecular gas, first observed in the CO J=2-1 line \citep{zuc95} and 
later confirmed in J=3-2 \citep{den05}.  At a distance of only 61 pc 
(Hipparcos), it is one of the closest known gas-rich circumstellar 
disks, farther only than TW Hydrae \citep[51pc;][]{mam05}.  Its outward 
similarity to a debris disk, combined with the substantial quantity of 
molecular gas still present in the system, suggest that the disk may be
in an unusual transitional evolutionary phase.

All three high-$\tau$ A stars are young: HR 4796A has an age of $8 \pm 2$ Myr \citep{sta95}
and $\beta$ Pic has been placed at $\sim 20$ Myr by \citet{thi01a}, consistent 
with the age determination of $20 \pm 10$ Myr by \citet{bar99}.  
The age of 49 Ceti is uncertain due to its isolation; unlike $\beta$ Pic or
HR 4796A there are no known associated low-mass stars to provide a corroborating
age estimate.  \citet{jur98} demonstrate that on an HR diagram, all three 
stars exhibit a low luminosity for their color, which is likely attributable 
to their young ages ($\sim 10$~Myr).  Using the evolutionary tracks of 
\citet{sie00}, \citet{thi01a} assign an age of 7.8~Myr to 49 Ceti based on 
its position on the HR diagram. 

Few conclusive measurements have been made of the dust properties in the 
49 Ceti system. HST/NICMOS coronographic observations of 49 Ceti failed to 
detect any scattered light in the near infrared at $r>1\farcs6$ \citep{wei99}. 
Recent subarcsecond-scale imaging at mid-infrared wavelengths with Keck 
\citep{wah07} revealed dust emission at 12.5 and 17.9 $\mu$m, extended along 
a NW-SE axis and apparently inclined at an angle of $60^\circ$. Simple 
models of the dust emission suggest a radial size segregation of 
dust grains, with a population of very small grains ($a \sim 0.1\mu$m) 
confined between 30 and 60 AU from the star, and a population of larger 
grains ($a \sim 15 \mu$m) from 60 to 900 AU from the star. 
However, the outer radius of this latter component is uncertain due to its
dependence on the millimeter flux, which is not well determined.  There are two 
contradictory single dish measurements of the millimeter dust emission, both 
with modest signal-to-noise.  \citet{boc94} report a IRAM 1.2~mm flux of 
$12.7 \pm 2.3$ mJy, while \citet{son04} report a JCMT/SCUBA 850 $\mu$m flux of 
$8.2 \pm 1.9$ mJy.  These measurements are mutually inconsistent for either
a thermal spectrum ($F_\lambda \propto \lambda^{-2}$) or a typical optically
thin circumstellar disk spectrum ($F_\lambda \propto \lambda^{-3}$) in this
wavelength regime.  

If we accept the lower value of the 850 $\mu$m flux and make standard 
assumptions about the dust opacity \citep[e.g.][]{bec91}, then the total mass 
of the 49 Ceti dust disk is 0.1 M$_\earth$.  If we compare this to other
nearby dusty disks at potentially similar stages of evolution, we find that
49 Ceti, with an 850 $\mu$m flux of 8.2 mJy at a distance of 61 pc, has a 
dust mass ($\propto F_{850\mu m} d^2$) approximately 80\% that of $\beta$ 
Pic \citep[104.3 mJy, 19.3 pc;][]{hol98} but only 0.3\% that of 
the typical Herbig Ae star HD 169142 \citep[554 mJy, 145 pc;][]{syl96}.
Thus the 49 Ceti disk appears to have a tenuous dust disk more akin to that
of the debris disk around $\beta$ Pic than a gas-rich protoplanetary disk.

Studies of the distribution of gas in the 49 Ceti system have been similarly 
inconclusive, particularly since it is not obvious that a substantial 
reservoir of molecular gas should persist in the strong UV field of an A star 
at this apparently advanced stage. Attempts to detect pure rotational 
transitions of the H$_2$ molecule have resulted in contradictory reports, 
with \citet{thi01n} reporting a marginal detection using SWS/ISO, which 
\citet{che06} did not confirm with Spitzer/IRS observations; nor did 
\citet{car07} detect H$_2$ emission with VLT/CRIRES observations.  Models of 
the double-peaked JCMT CO J=3-2 line profile observed by \citet{den05} 
indicated that the gas is likely distributed in either a very compact disk 
with $\sim 16^\circ$ inclination or a more inclined ring of radius $\sim 50$ AU 
and inclination $\sim 35^\circ$.  The latter was deemed more consistent with 
the dust distribution seen in the mid-infrared, although it fails to reproduce 
the high-velocity wings that may be present in the CO J=3-2 line profile. 

In order to obtain spatially resolved information on the distribution of  
material in the system, we observed 49 Ceti with the Submillimeter 
Array\footnote{The Submillimeter Array is a joint project between the 
Smithsonian Astrophysical Observatory and the Academia Sinica Institute of 
Astronomy and Astrophysics and is funded by the Smithsonian Institution and the 
Academia Sinica.}
in the J=2-1 transition of CO and associated continuum.  We detect a rotating 
structure of much greater extent than predicted from the single-dish 
measurements, with a large central region devoid of molecular gas emission.  
We also model the disk emission using COSTAR 
\citep{kam00,kam01}, a code that combines thin hydrostatic equilibrium 
models of disks with a rich chemistry network and a detailed heating and 
cooling balance to determine gas properties.  The models provide some insight 
into basic properties of the disk, including the region of photodissociation 
of CO in the inner disk and the spatial extent of the emission.  

The observations are described in \S\ref{sec:obs}, and results presented
in \S\ref{sec:res}.  In \S\ref{sec:mod} we discuss the process undertaken
to model the data, including the basic model structure, the initial conditions
for the chemistry, and the initial model adopted from the dust emission analysis
of \citet{wah07}, as well as adjustments to that fiducial model necessitated
by the new observations.  The parameter space is explored in 
\S\ref{sec:grid}, including an investigation of the varying influence of 
chemistry across the model grid, and \S\ref{sec:SED} discusses the dust
properties in the context of the spectral energy distribution predicted from
the gas model.  The best-fit model is discussed in \S\ref{sec:best}, including 
an a posteriori comparison of the model prediction with the observed CO J=3-2 
spectrum; inadequacies of the model are also noted.  The results are discussed
in the broader context of disk evolution in \S\ref{sec:dis}, and a summary is 
presented in \S\ref{sec:con}.

\section{Observations}
\label{sec:obs}

We observed 49 Ceti with the SMA at 230 GHz during an 11-hour track on
the night of October 13, 2006.	Atmospheric phase was extremely stable,
with typical phase changes of $<15^\circ$ between calibrator scans (every 25
minutes).  Seven antennas were used in the ``extended'' configuration, with 
projected baselines between 15 and 130 meters.
The primary flux calibrator was Uranus, and the passband calibrators were
the quasars 3C454.3 and J0530+135. Gain calibration was carried out using
the quasar J0132-169, located just 1.3$^\circ$ from 49 Ceti; the flux derived
for this quasar was 0.93 Jy.  The nearby quasar J0006-063 was also included
to test the quality of the phase transfer from J0132-169.

Two 2-GHz sidebands separated by 10 GHz were used, yielding a continuum
sensitivity of 0.7 mJy (1$\sigma$).  Spectral resolution in the line was
0.26 km/s, subsequently binned to 2.1 km/s, with rms sensitivity 0.030 Jy in a 
single 2.1 km/s channel.  The LSR velocities were converted to heliocentric 
using an offset of -9.14 km/s.  The synthesized naturally weighted beam in the 
CO J=2-1 line was 1\farcs0$\times$1\farcs2, at a position angle of 
-78.6$^\circ$.
Imaging was carried out using the MIRIAD software package.

\begin{figure}
\centering
\includegraphics[scale=0.6,angle=90]{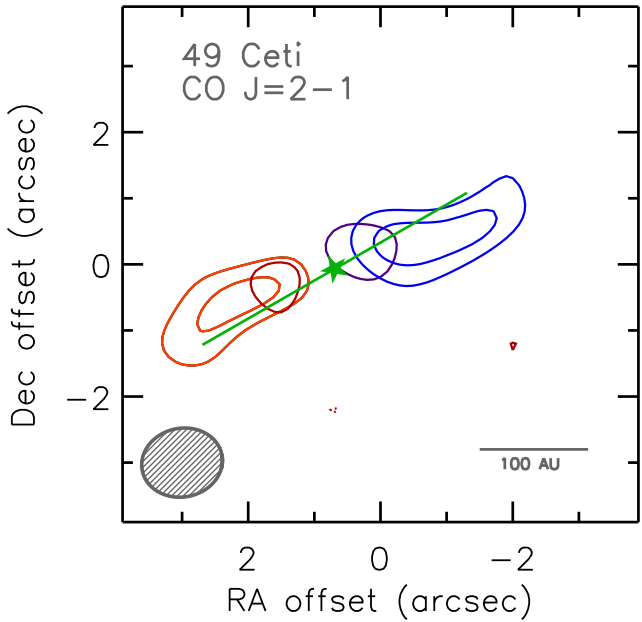}
\caption{A renzogram of SMA observations of 49 Ceti in the CO J=2-1 line.  
The beam size is 1\farcs0$\times$1\farcs2, and the position angle is 
$-79^\circ$.  Contours are -3, 3, and 5 $\times$ 37 mJy/beam (the rms noise).  
The position of 49 Ceti is marked with a star symbol, while the green line 
indicates the position angle derived by \citet{wah07} from mid-IR imaging.
The contour colors indicate heliocentric line-of-sight velocity; 
the four distinct velocities shown are 9.0, 11.1, 13.2, and 15.3 km/s, in the
order of bluest to reddest channel.  No emission was detected outside this
velocity range.
}
\label{fig:map}
\end{figure}

\section{Results and Analysis}
\label{sec:res}

Figure \ref{fig:map} shows the observed line emission from the region around
49 Ceti.  Four velocity channels are shown, with the velocity indicated by 
the color of the contour lines.  The observations are centered on the J2000
coordinates of 49 Ceti; the star symbol indicates the position corrected
for the proper motion measured by {\em Hipparcos}.  The maximum signal-to-noise 
ratio in the line is 8.  The CO J=2-1 emission appears to be in an extended 
rotating structure of $>2$" radius, apparently viewed close to edge-on.  
The symmetric distribution of the emission in the four velocity channels 
implies a heliocentric velocity near 12.2 km/s, consistent with previous 
determinations of the systemic velocity \citep[10.5 and 9.9 km/s for the disk 
and the star, respectively; see][and references therein]{den05}. 
No emission is detected outside the range of velocities shown.
The wide separation of the emission peaks, combined with a lack
of compact, high-velocity emission, suggests that the central regions are
clear of CO J=2-1 emission out to $\sim$90~AU radius ($\sim$1\farcs5),
twice that of the larger ring predicted from the CO J=3-2 single dish data.
Table \ref{tab} lists the observed parameters of the system.

\begin{table*}
\caption{Observational parameters for 49 Ceti}
\begin{center}
\begin{tabular}{lcccc}
\hline
\multicolumn{1}{c}{Parameter} & $^{12}$CO(3-2)$^a$ & $^{12}$CO(2-1) & $^{13}$CO(2-1) & continuum \\
\hline
\hline
Rest frequency (GHz) & 345.796 & 230.538 & 220.399 & 230.5 (USB$^b$) \\
Channel width & 0.27 km s$^{-1}$ & 2.1 km s$^{-1}$ & 8.4 km s$^{-1}$ & 2$\times$2 GHz \\
Beam size (FWHM) & 14" & 1\farcs0$\times$1\farcs2 & 1\farcs0$\times$1\farcs2 & 1\farcs0$\times$1\farcs2 \\
~~~PA & -- & -78.6$^\circ$ & -78.6$^\circ$ & -78.6$^\circ$ \\
rms noise (Jy beam$^{-1}$) & 0.22 & 0.030 & 0.017 & $7.0\times 10^{-4}$ \\
Dust flux (mJy) & -- & -- & -- & $<2.1$ \\
Peak brightness temperature (K) & 0.076$\pm$0.008 & 3.5$\pm$0.5 & $<0.8$ & -- \\
Integrated intensity (Jy km s$^{-1}$) & 9.5$\pm$1.9 & 2.0$\pm$0.3 & $<0.5$ & --
\\
\hline
\end{tabular}
\end{center}
\tablenotetext{a}{~\citet{den05}}
\tablenotetext{b}{~Upper sideband frequency; lower sideband is centered at 
220.5 GHz.  Both sidebands have 2 GHz width.}
\label{tab}
\end{table*}

Assuming optically thin lines and LTE, the total mass in CO probed by the
J=2-1 transition is given by
\begin{eqnarray}
M=\frac{4 \pi}{h \nu_{21}} \frac{F_{21} m d^2}{A_{21} x_{2}}
\end{eqnarray}
where the subscript 21 refers to the CO(2-1) transition, $F$ is the
integrated flux in the line, $d$ is the distance to the source (61 pc;
Hipparcos), $m$ is the mass of the CO molecule, $\nu$ is the rest frequency
of the transition, $h$ is Planck's constant, and
$x_{2} \equiv \frac{N_2}{N_{tot}}$ where $N_2$ is the population in the
J=2 rotational level while $N_{tot}$ is the total CO population.
The CO mass calculated using this method is $2.2 \times 10^{-4}$ M$_\earth$.
Using the canonical CO/H$_2$ ratio of 10$^{-4}$ this yields a molecular
hydrogen mass of $2.2$ M$_\earth$, consistent with the value
of $6.3\times 10^{-3}$ M$_{Jup}$ = $2.0$ M$_\earth$ calculated by \citet{zuc95}.

No continuum emission was detected at this combination of resolution and
sensitivity.  This indicates one of two things: either the continuum flux is
concentrated at the center of the disk but the total flux is too low to be
detected, or the total flux may be larger but spread over many beams, so that 
the brightness within each beam is below our detection threshold.  These 
observations were sensitive enough to detect the higher continuum flux 
reported by \citet{boc94} if it were concentrated within a few synthesized 
beams.  However, an extrapolation of the \citet{son04} value for a typical 
circumstellar dust spectrum predicts a lower flux by a factor of 6, which is 
just below the detection threshold.  The lack of an SMA continuum detection 
at 230\,GHz is therefore inconclusive: if the \citet{son04} value is correct, 
we would not expect to detect even centrally concentrated emission, and so we 
cannot constrain the spatial extent of dust emission through the nondetection 
at 230\,GHz.

\section{Disk Modeling}
\label{sec:mod}

In order to gain insight into the physical processes at work in the 49 Ceti
system, we carried out modeling of the disk with COSTAR \citep{kam00,kam01},
a code which solves the chemical equilibrium simultaneously with a detailed 
heating and cooling balance to determine gas properties of circumstellar disks.
In the following, the salient features of these models are summarized. 
The chemistry is modeled using a network of 48 different species covering the 
elements H, He, C, O, S, Mg, Si, and Fe. The elemental abundances and key 
parameters of these models, including the stellar mass, radius, effective 
temperature, surface gravity, and ultraviolet flux, are summarized in 
Table~\ref{models_par}. The 48 species are connected through 281 reactions, 
including cosmic ray chemistry, photochemistry and the chemistry of excited 
H$_2$. We compute equilibrium chemistry using a modified Newton-Raphson 
algorithm. The solution then only depends on the element abundances and not 
on initial conditions.

We use the results of dust modeling by \citet{wah07} and assume large 
$30~\mu$m black body grains with radiative efficiencies of $Q_\lambda = 2\pi 
a/\lambda$ for $\lambda > 2\pi a$ and $Q_\lambda = 1$ otherwise. These 
grains are efficient absorbers and inefficient emitters, thus achieving dust 
radiative equilibrium temperatures of \begin{equation}
T_{\rm dust} = 324 \left(\frac{L_*}{L_\odot}\right)^{0.2} (a_{\rm \mu m})^{-0.2} (r_{\rm AU})^{-0.4}~~~{\rm K}\,\,\,.
\end{equation}
Here, $L_*$ and $L_\odot$ are the stellar and solar luminosity 
respectively, $a_{\rm \mu m}$ is the grain size in micron and $r_{\rm AU}$ the 
distance from the star in astronomical units.  The gas temperature is derived 
from a detailed energy balance including the most relevant heating and cooling 
processes \citep{kam01}.  

\begin{table}[htdp]
\caption{Element abundances and parameters used in the disk models}
\begin{center}
\begin{tabular}{lc}
\hline
Parameter$^a$      & Value \\
\hline
\hline
$A_{\rm He}$   &  $7.5 \times 10^{-2}$ \\ 
$A_{\rm C}$    &  $1.3 \times 10^{-4}$ \\ 
$A_{\rm O}$    &  $2.9 \times 10^{-4}$ \\ 
$A_{\rm Mg}$   &  $4.2 \times 10^{-6}$ \\
$A_{\rm Si}$   &  $8.0 \times 10^{-6}$ \\
$A_{\rm S}$    &  $1.9 \times 10^{-6}$ \\
$A_{\rm Fe}$   &  $4.3 \times 10^{-6}$ \\
T$_{\rm eff}$  &  10\,000~K            \\
$\log g$       & 4.5                   \\
R$_\ast$       & 1.7~R$_\odot$         \\
M$_\ast$       & 2.3~M$_\odot$         \\
$\sigma_{\rm UV}$ & $4.68\,10^{-24}~{\rm cm}^{-2}~{\rm H-atom}^{-1}$\\
\hline
\end{tabular}
\tablenotetext{a}{Gas-phase abundances ($A$) are relative to hydrogen.}
\end{center}
\label{models_par}
\end{table}

\begin{center}
\begin{figure}[htp]
\centering
\includegraphics[scale=0.6,viewport=110 30 400 350]{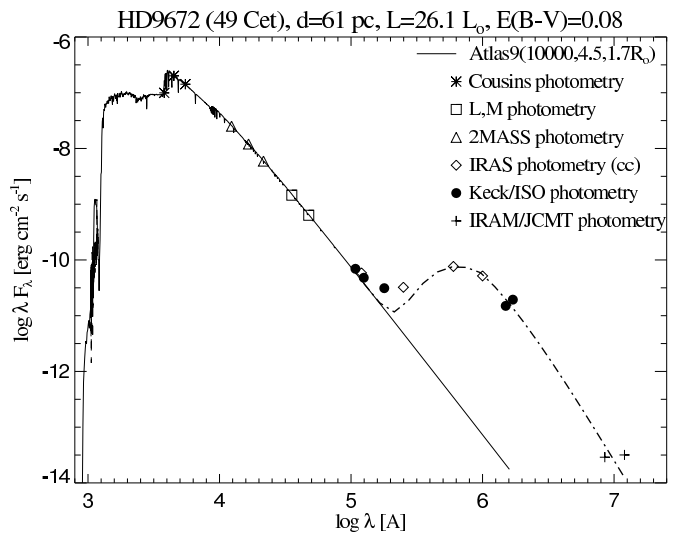}
\caption{
Spectral energy distribution (de-reddened according to extinction derived 
by \citealt{syl96} and \citealt{car89} extinction law) for 49 Ceti using 
available optical, infrared, and submillimeter photometry.  The solid line
denotes a Kurucz stellar atmosphere model fitted to the photometry using
the Hipparcos distance of 61 pc. 
The dot-dashed line shows the SED for the best-fit model of the outer disk 
see text of \S\ref{sec:SED} for details.
}
\label{fig:SED}
\end{figure}
\end{center}

The radiation field consists of both stellar and interstellar components. 
The stellar properties are determined by a Kurucz model fit to photometric
points collected from the literature \citep{wah07,syl96,boc94,son04}; 
using T$_{\rm eff}$=10000~K and $\log g$ = 4.5,
consistent with the values quoted by \citet{che06}, the derived stellar 
luminosity is $L_*=26.1 L_\sun$ and the radius is 1.7 R$_\sun$. 
The spectral energy distribution and Kurucz model are plotted in Figure
\ref{fig:SED}, including dereddening according to extinction derived
by \citet{syl96} and using a \citet{car89} extinction law.  The solid
line in the figure denotes the fit to the photometry of a Kurucz stellar
atmosphere model at the Hipparcos distance of 61 pc.  The dashed line shows
the spectral energy distribution of the best-fit model of the outer disk as 
described in \S\ref{sec:best}. 
The interstellar radiation field in the ultraviolet is assumed to be 
$1.2 \times 10^7$ cm$^{-2}$ s$^{-1}$ \citep{hab68}.  

A basic model of the dust disk was constructed according to the 
Bayesian analysis of mid-infrared emission carried out by \citet{wah07}.  
Their model consists of an inner disk extending from 30 to 60 AU, composed 
primarily of small grains ($a \sim 0.1$ $\mu$m) with a surface density of 
$5\times10^{-8}$ g/cm$^2$, and an outer disk extending from 60 to 900 AU 
composed of larger grains ($a \sim 15$ $\mu$m) with a surface density of 
$3 \times 10^{-6}$ g/cm$^2$.  They derive a surface density distribution for 
the outer disk that is constant with radius, yielding a total disk mass of 
0.35 M$_\earth$.  From the mid-IR images, they also determine a position angle 
of $125^\circ \pm 10^\circ$ (indicated in Figure \ref{fig:map}) and an 
inclination of $60^\circ \pm 15^\circ$.  We use this model as a starting
point for the disk structure, since it reflects the best available information
on the dust density distribution.  However, since the molecular gas emission
provides better constraints on some aspects of disk structure, including the 
vertical density distribution and the surface density structure of the outer 
disk, we introduce refinements to this initial model where justified, as 
described in \S\ref{sec:rad} and \S\ref{sec:grid} below.  For the large grain
population, our model uses 30~$\mu$m grains instead of 15~$\mu$m grains, 
although the grain size used in this simple model is highly degenerate with 
other disk properties, as discussed in \S\ref{sec:SED}.

To predict gas properties from this dust model, we make two 
primary assumptions: (1) gas and dust are well-mixed, (2) the gas:dust mass
ratio is constant.  We initially assume a constant scale height $H$=2\,AU,
since there is no information on disk scale height from the dust model of 
\citet{wah07}; we also begin by retaining the inner and outer radii and 
radially constant surface density structure from the \citet{wah07} model, 
although these assumptions are modified in \S\ref{sec:rad} below.
Throughout the modeling process, we use the canonical gas:dust mass ratio of 
100 and  assume that the disk is embedded in interstellar material of density 
10~cm$^{-3}$ to avoid model densities dropping to unrealistically low values 
near the boundaries of the numerical grid.

To compare our models with the SMA data, we use the radiative transfer 
code RATRAN \citep{hog00} to generate a sky-projected image of the CO J=2-1 
emission predicted for the physical model.  We then use the MIRIAD task {\em 
uvmodel} to sample the image with the combination of spatial frequencies 
and visibility weights appropriate for our SMA data.  We allow the inclination 
and position angle of the system to vary in order to best match the data.

\subsubsection{Inner Disk}

In the inner disk, inside 60 AU, composed primarily of small grains, 
the stellar radiation field raises the dust temperature to 1000-2000 K and 
dissociates most of the molecular gas.  In this region, the dominant form of 
carbon is C$^+$, and even hydrogen is predominantly atomic.  We therefore 
ignore the inner disk component in subsequent modeling and focus on 
reproducing the observed CO emission with only the outer disk component.  

This lack of molecular gas in the inner disk is consistent with the 
non-detection of warm H$_2$ by \citet{che06} and \citet{car07}, and with the 
lack of high-velocity CO emission in Figure \ref{fig:map}.
The lack of CO emission more than 4.3 km/s from the stellar velocity is 
consistent with an absence of CO within a radius of $\sim90$ AU, 
for gas in Keplerian rotation around a star of 2.3~$M_{\sun}$.

\subsubsection{Outer Disk}
\label{sec:rad}

There are three primary features of the observed CO emission from the outer 
disk that we attempted to reproduce with this modeling effort: (1) the 
separation of the emission peaks in the outer channels ($\sim3$"), (2) the 
spatial extent of the CO emission in all channels, and (3) the strength of the 
emission. Reproducing these features of the observed CO emission requires 
several modifications to the best-fit \citet{wah07} model of the outer 
dust disk, including adjustments to the inner and outer radii and a departure
from the constant surface density prescription. 

At first glance, the inner radius of 60 AU derived by \citet{wah07} might 
seem consistent with the lack of emission within 90 AU derived from the 
missing high-velocity wings in our data; however, there is a large region at 
the inner edge of the outer disk subject to photodissociation by stellar 
radiation which therefore contributes little to the CO emission.  In order to 
reproduce the separation of the emission peaks, material is required 
interior to this 60 AU radius.  We therefore take the uncertainties in the 
\citet{wah07} dust distribution into account and allow the inner disk radius to 
vary.  However, moving the inner radius closer than $\sim 40$ AU to the star 
results in high-velocity emission that we do not observe in the data, while 
still producing emission peaks wider than observed.  We therefore set the
disk inner radius at 40 AU, and then adjust the gas densities to further 
reduce the separation of the emission peaks.

Increasing the total gas mass leads to an elongated morphology with an aspect 
ratio larger than the observations, as the optical depth rises throughout the 
disk.  To meet the three criteria of (1) enough gas-phase CO near the inner 
disk edge to reproduce the observed peak separation, (2) low enough optical 
depth in the outer parts of the disk to keep the emission from becoming more
elongated than the data (through photodissociation by interstellar UV 
photons), and (3) maintaining an inner radius large enough to avoid generating 
high-velocity emission that is not present in the data, we must ``pile up'' 
material at the inner disk edge to enhance shielding and concentrate 
emission.  We therefore modify the initial assumption of constant surface 
density as derived from the infrared analysis, instead adopting an 
$r^{-\epsilon}$ density profile.  We simultaneously relax the constant scale
height assumption, introducing a scale height $H$ that increases linearly 
with radius $r$, with proportionality constant $h=H/r$. The full 2-D density 
structure then becomes $n(r,z) = r^{-\epsilon} \exp{(-z^2/2 H^2)}$, where the 
exponent $\epsilon$ and scale height constant $h$ are varied to obtain the 
best fit to the CO data.  

The power-law surface density profile results in a much better match between 
the model and the observed emission peak separation.  It also curbs the 
elongation of the emission to some extent, as the vertical column density of 
the outer disk drops and the material far from the star becomes subject to 
dissociation by interstellar radiation. However, even steep power law indices 
for the surface density profile do not result in a completely photoevaporated 
outer disk and consequently produce emission that is much more elongated than 
observed.  In a next step, we therefore reduce the outer radius from 900 to 
200~AU.  While this is at the lower end of the range allowed by \citet{wah07}, 
their derived outer radius was based largely on the uncertain millimeter flux 
measurement, and the gas geometry is likely a better probe of the disk extent.

\subsection{Grid of Disk Models}
\label{sec:grid}

After these initial studies of the outer disk, it became clear that several 
model parameters were ill-constrained by previously existing data.
Specifically, the disk mass is constrained only by the weakly-detected and 
contradictory millimeter flux measurements; similarly, the density power law 
index $\epsilon$ is ill-determined by the infrared observations, which are 
primarily sensitive to inner disk emission. The scale height $h$ is also 
completely unconstrained by the continuum or single-dish measurements, 
neither of which is sensitive to disk structure in the vertical direction.  
The disk geometry (PA and inclination) quoted by \citet{wah07} is also 
subject to large uncertainties, due to the irregular shape of the emission 
observed in the infrared.  We therefore attempt to better constrain these 
disk parameters by using our resolved CO gas line observations. Gas lines 
are generally more sensitive than dust emission to temperature and density 
gradients, and can thus provide means to break model degeneracies. We 
ran grids of models for the three structural parameters (disk mass, 
density index, scale height) and two geometrical parameters (PA, 
inclination), finding the best-fit values by calculating and minimizing 
a $\chi^2$ value comparing the model to the observed emission from the disk.  
Due to the computational intensity of the calculations necessary to 
determine the chemistry and radiative transfer solutions for each model, 
we ran only a sparsely sampled grid of models.  In order to ensure that the
final model reflects all available observational constraints, we centered the 
grid on the fiducial model of \S\ref{sec:mod} and adjusted the parameters only 
as necessary to better reproduce the new CO(2-1) observations, moving from 
coarse to fine grids to ensure adequate exploration of the parameter space.  
We use the modeling primarily as a demonstration that the basic features of 
the observed CO emission can be reproduced by a simple azimuthally symmetric 
model of disk structure; the ``best-fit'' model should therefore be viewed 
as representative of an initial understanding of the features of the 
system rather than as a conclusive determination of the disk structural 
parameters.

\subsubsection{CO Chemistry Across the Model Grid}

The CO chemistry is dominated by photodissociation in a number of UV bands
and thus the abundance of CO in each model is mostly dependent on the radial
and vertical column densities being able to shield the stellar and interstellar
UV radiation respectively. In the following we briefly discuss some basic 
characteristics of the model grid.

The surface density in the models is independent of the scale height and hence
the radial mass distribution in each model can be written as $M(R) \propto R^{-
\epsilon +3}$, where $M(R)$ denotes the mass inside a radius $R$. So, as we
increase the density power law exponent $\epsilon$, the inner region of the 
outer disk harbors a larger fraction of the total mass. The densities in this 
region of the disk become higher and hence it is easier to obtain the critical 
column densities necessary for UV shielding in the radial direction. On the 
other hand, a shallower gradient for the density distribution 
translates into higher densities in the outer parts of the disk, thus 
enhancing the vertical shielding in the outer disk compared to models with 
high $\epsilon$.  None of our models is optically thick in the dust continuum, 
so the UV shielding is mainly H$_2$ shielding of the CO bands due to their 
overlap in wavelengths; CO self-shielding also plays a role.

\begin{figure*}
\centering
\includegraphics[width=0.99\textwidth]{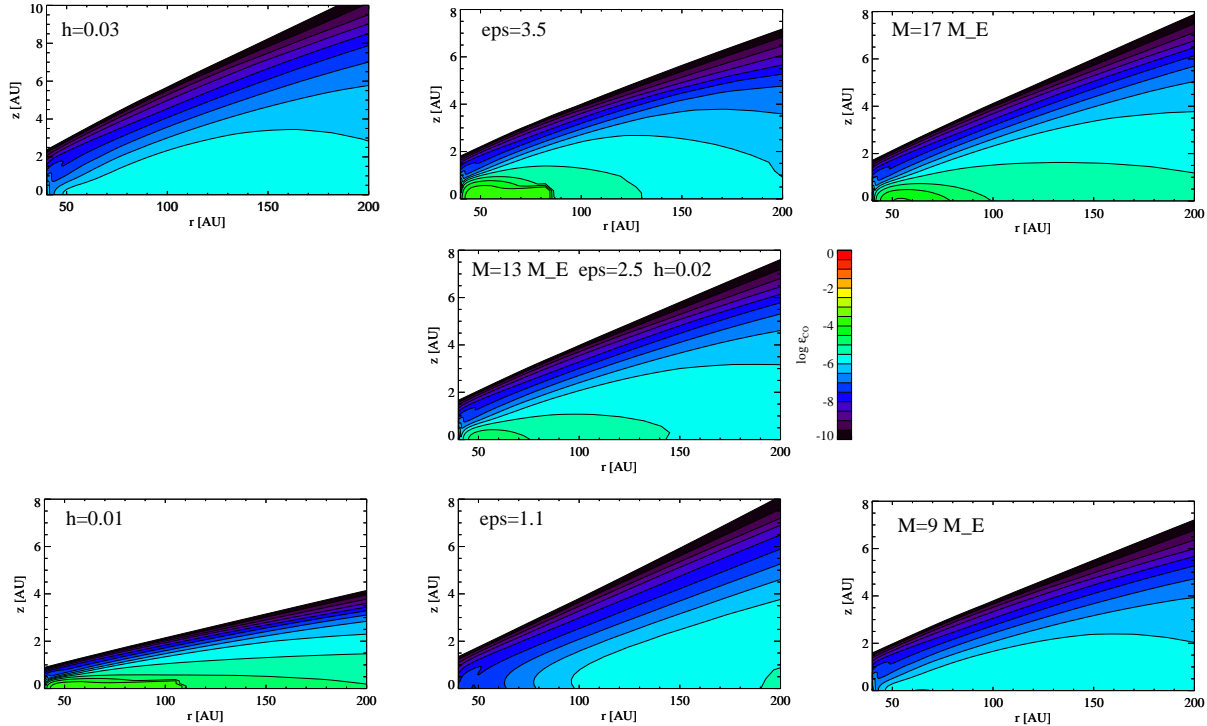}
\caption{
Two-dimensional CO abundances in a subset of disk models.  The center panel 
shows the best-fit model ($M=13$M$_\earth$, $\epsilon = 2.5$, $h=0.02$), while 
the rows of models above and below show the effects of incrementing and 
decrementing, respectively, each of the three structural parameters that were 
allowed to vary during the fitting process: $h$ {\em (left column)}, $\epsilon$ 
{\em (center column)}, and M$_{disk}$ {\em (right column)}.  The values of the
parameters shown are $h= 0.01$, $0.03$; $\epsilon=1.1$, $3.5$; and $M=9$, $17$ 
M$_\earth$.  
}
\label{COoverview1}
\end{figure*}

With this basic picture, we can understand the CO chemistry displayed in Fig.~
\ref{COoverview1} as a function of disk mass (right column) and density 
gradient $\epsilon$ (center column).
As the total disk mass is increased, CO first starts to build up in the radial 
direction.  It can still be dissociated by the vertically impinging 
interstellar UV radiation field in the outer regions of the disk (150-200~AU) 
until the disk reaches a mass of $\sim17$~M$_{\earth}$, at which point it 
becomes opaque in the CO bands even in the vertical direction. A shallow 
density gradient always leads to smaller radial column densities at the same 
reference radius, thus pushing the C$^+$/C/CO transition further out in radial 
distance. In our best-fit model of 13~M$_{\earth}$, a change in $\epsilon$ 
from 2.5 to 1.1 changes the radius for the C$^+$/C/CO front from close to 
40~AU to 190~AU. 

The scale height $h$ of the models affects only the vertical density structure 
in the models. However, since density and chemistry are closely intertwined, it 
can strongly impact the overall radial and vertical structure of the CO 
chemistry.  From a comparison of the center panel with the bottom left panel in
Figure \ref{COoverview1}, we see that a factor 2 lower scale height with 
respect to the best fit model ($h=0.02$), enhances the CO abundance in the disk 
significantly, leading to radial and vertical column densities that are more 
than a factor 10 higher with respect to the best fit model. The total CO mass 
increases by a factor of 10 as well, with the integrated emission undergoing
a corresponding dramatic increase.  

Table~\ref{grid_CO_tab} displays some key results from a subset of grid models
such as characteristic radial and vertical CO column densities, CO masses and
total CO J=2-1 line emission.  For all models in the table, the inner radius 
is fixed at 40~AU and the outer radius at 200~AU.  

\begin{table*}[htdp]
\caption{Derived quantities from a subset of the 49 Ceti disk models}
\begin{center}
\begin{tabular}{rccrrrr}
\hline
$M_{\rm disk}^a$ & $\epsilon$ & $h$ & $N({\rm CO})_{\rm radial}^b$ & $N({\rm
CO})_{\rm vertical}^{\rm 100 AU,c}$  & $M_{\rm CO}^d$ & $I_{\rm CO}$(J=2-1)$^e$ \\
 (M$_{\earth})$  &                      &               &  
($10^{18}$\,cm$^{-2}$)     &  
($10^ {15}$\,cm$^{-2}$)
 & ($10^{-4}$\,M$_{\earth}$)    & (Jy km s$^{-1}$) \\
\hline
\hline
13 & 2.5 & 0.020 &   2.76 &  4.23  &  9.66 &  2.6 \\
 9 & 2.5 & 0.020 &   0.32 &  1.82  &  2.46 &  1.2 \\    
17 & 2.5 & 0.020 &   13.5 &  9.06  &  37.2 &  6.9 \\
13 & 3.5 & 0.020 &   15.1 &  4.47  &  98.0 & 11.7 \\
13 & 1.1 & 0.020 &   0.13 &  0.91  &  3.74 &  2.3 \\
13 & 2.5 & 0.010 &   42.8 &  78.4  &  96.6 & 14.5 \\
13 & 2.5 & 0.030 &   0.12 &  2.20  &  2.97 &  1.5 \\
\hline
\end{tabular}
\tablenotetext{a}{Total disk gas mass}
\tablenotetext{b}{Total radial CO column density through the midplane}
\tablenotetext{c}{CO vertical column density at 100 AU}
\tablenotetext{d}{Total CO mass in the disk}
\tablenotetext{e}{Integrated CO(J=2-1) line emission}
\end{center}
\label{grid_CO_tab}
\end{table*}

\subsubsection{From Chemistry to Observables}

\begin{figure*}
\centering
\includegraphics[angle=90,totalheight=0.66\textheight]{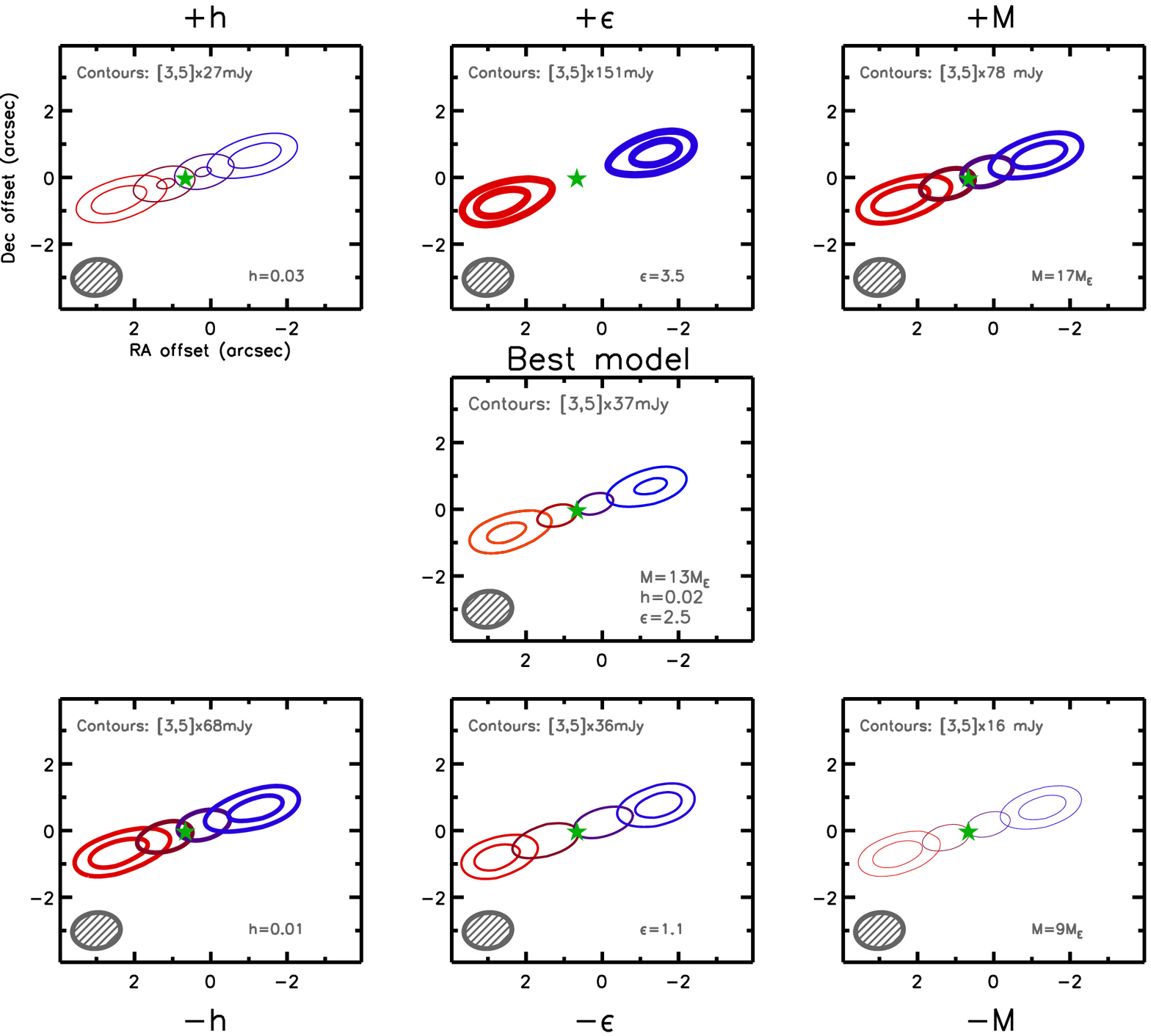}
\caption{
CO J=2-1 emission predicted for the subset of models shown in Figure 
\ref{COoverview1}, sampled with the same spatial frequencies and visibility 
weights as the SMA data in Figure~\ref{fig:map}.  The center panel shows the 
best-fit model, while the rows of models above and below show the effects of 
incrementing and decrementing, respectively, each of the three structural 
parameters that we allowed to vary during the fitting process: $h$ {\em (left 
column)}, $\epsilon$ {\em (center column)}, and M$_{disk}$ 
{\em (right column)}.  The contour levels are displayed in the upper left 
corner of each panel; they are set at 3 and 5 $\times$ 15\% of the peak flux 
for each model. The thickness of the contours is proportional to the absolute 
flux: thicker contours indicate that the source is brighter than the data, 
while thinner contours indicate that it is fainter than the data.  The contour
levels in the center panel are identical to those in Figure~\ref{fig:map}.  
Table \ref{tab:best} gives the full list of parameters for the best-fit model.  
}
\label{fig:model}
\end{figure*}

The predicted CO J=2-1 emission for the models in Figure~\ref{COoverview1} 
is displayed in Figure~\ref{fig:model}; a comparison of these figures
illustrates the ways in which differences in chemical structure are manifested
in the observable properties of the CO emission.  The CO emission is sampled
with the same spatial frequencies and visibility weights as the SMA data and 
displayed in renzogram form with the same velocity structure as in 
Figure~\ref{fig:map}.  In order to emphasize the relative structural 
differences between models, the contour levels are 15\% of the peak flux for 
each model, with the absolute flux indicated by the thickness of the 
contours, and also printed explicitly at the top of each panel.  

The decreased shielding in the inner disk caused by reducing the density 
gradient $\epsilon$ is visible as a lengthening of the emission in the central
channels and a widening of the emission peaks in the outer channels in the 
low-epsilon model (bottom center panel).  Increasing $\epsilon$ (top center 
panel) leads to enhanced shielding at the disk inner edge, causing much higher 
CO fluxes in the outer part of the disk and extremely high contrast between the 
inner and outer velocity channels. 

The primary observable consequence of adjusting the mass (right panels, top 
and bottom) is that the increased or decreased shielding from extra gas leads 
to a corresponding increase or decrease in the total CO flux; changes to the 
shape of the emission are minimal, and the primary difference between models 
of different mass over the mass range under consideration is simply in the 
relative brightness of the emission.  

Differences in the scale height of the disk similarly manifest as differences
in the flux scale; however, decreasing the scale height (bottom left panel) 
also causes greater shielding at the inner disk edge, leading to greater 
elongation of the outer velocity channels and causing the inner velocity 
channels to draw together and overlap as the CO flux rises throughout the 
inner areas of the disk.  An increase in scale height (top left panel) leads 
to a greater area in the front and back of the disk, projected along 
our line of sight, which increases the flux in the central channels and leads 
to a lower contrast between the inner and outer channels of the disk.

\subsection{Spectral Energy Distribution}
\label{sec:SED}

After converging initially on a model that was able to reproduce the observed
CO J=2-1 emission, we used that model to predict the spectral energy 
distribution.  This serves as an {\em a posteriori} test of the consistency 
between the gas and dust properties in the models and the available observables.

We integrate over the disk volume to obtain the flux as a function of 
wavelength 
\begin{equation}
F_\lambda = (\pi a^2 / d^2) \int \int 2 \pi r \, B_\lambda(T_{dust}(r,z)) n_{dust}(r,z) \, Q(\lambda) \, dz \, dr\,\,\,, 
\end{equation}
where $d$ is the distance to the source and $n_{dust}$ is the number density of 
dust grains in cm$^{-3}$.  We assume throughout a grain density of 2.5 g/cm$^3$.

While the predicted shape of the spectral energy distribution matches the
observations well, the absolute fluxes are too high by a factor of $\sim 5$.  
Adjusting the temperature of the dust grains alters the shape of the SED curve, 
causing it to deviate from the observed shape; we were therefore required to 
increase the gas:dust ratio from 100 to 500 in order to reproduce the observed 
photometry.  This unusually high ratio is likely an artifact of the simple 
assumptions of the model, since little information is available about the dust
distribution in this system (and none at all from our data).  For example, the 
mass of the system is likely not all in 30 $\mu$m grains, and a significant 
fraction of the mass may be in larger grains that contribute little to the 
infrared emission.  Another possibility is that the overall gas:dust ratio is
consistent with the standard value, but that gas and dust are not well-mixed:
for example, much of the excess emission may arise from the inner edge of the
disk, which will be directly illuminated and heated by the stellar radiation.
Resolved observations of the dust continuum emission would test this hypothesis
by placing constraints on the spatial distribution of the emitting region.
Including effects such as this would significantly complicate the model
presented here, as the H$_2$ formation rate would be affected by varying the
abundance of the dust on which it forms.  In general, the dust size and 
gas:dust mass ratio are strongly related by the total dust surface required 
to maintain the observed quantity of molecular gas; these are in turn 
dependent on the stellar properties determining dust grain temperatures.  None
of these dust-dependent quantities are well constrained by available data.  
Given the observations available and the extremely simplified dust model, 
which not only neglects the size distribution but also the possibility of a 
mixture of compositions and opacities, we use the simplest assumption of an 
altered gas:dust ratio in order to conduct a consistency check of the 
temperature and density structure of the gas model.

Decreasing the total dust mass in the model to match the SED reduces the grain 
surface area for H$_2$ formation. Thus molecular hydrogen begins to form at 
larger radii and greater depth, compared to the initial model with the 
canonical gas:dust ratio of 100.  As a consequence of less effective UV 
shielding, the total CO mass decreases. Hence the total mass of the best-fit 
model has to be increased slightly to compensate for the lower molecular gas 
fraction. As a secondary effect, the overall gas temperature of the 
dust-depleted model also decreases due to the diminished photoelectric 
heating in the disk.  The corresponding SED predicted for these parameters 
is indicated by the dashed line in Figure \ref{fig:SED}.  The mid-infrared 
flux points are underestimated by this model because we do not include the 
inner disk component of \citet{wah07}; as our data provide no constraints 
on the properties of the inner disk, we ignore this component and 
concentrate on the fit to the outer disk.  The flux predicted by the model 
SED is consistent with our own continuum upper limit reported in 
Table~\ref{tab}.

\subsection{Best-Fit Disk Model}
\label{sec:best}

The center panel of Figure~\ref{fig:model} shows the best-fit model from the 
grid, with the minor modifications introduced by reproducing simultaneously the 
spectral energy distribution.  The structural and geometric parameters for this 
model are listed in Table \ref{tab:best}.  The errors given in the table are
the approximate 1-$\sigma$ uncertainty range interpolated from the $\chi^2$ 
grid.

\begin{table}
\caption{Parameters for Best-Fit Disk Model}
\begin{center}
\begin{tabular}{lc}
\hline
$h$ & $0.020^{+0.015}_{-0.005}$ \\
$\epsilon$ & $2.5^{+0.5}_{-1.0}$ \\
$M_{gas}$ & $13 \pm 3$ M$_\earth$ \\
$M_{dust}$ & $0.02 \pm 0.01$ M$_\earth$ \\
$i$ & $90^\circ \pm 5^\circ$ \\
PA & $-70^\circ \pm 10^\circ$ \\
$R_{in}$ & 40 AU$^a$ \\
$R_{out}$ & 200 AU$^a$ \\
\hline
\end{tabular}
\end{center}
\tablenotetext{a}{For a description of the constraints on the inner and
outer radii, see \S\ref{sec:rad}}
\label{tab:best}
\end{table}

This model reproduces the basic features of the CO J=2-1 emission well,
including the strength of the emission, the separation of the emission peaks,
and the spatial extent of the emission.  There are still several important
differences between the model and the data, however, including (1) an inability
to reproduce the changes in position angle with radius evident in the data
(the ``wings'' of emission extending to the southeast and northwest of the
position angle axis), and (2) the separation of the innermost, low-velocity 
channels.  Both of these may be indicative of departures from azimuthal 
symmetry in the disk structure, the former possibly indicating a warp in the
disk and the latter apparently pointing to a deficit of emission along the 
minor axis of the disk.  In none of our models were we able to reproduce the 
wide separation between the inner channels; while the signal-to-noise ratio in 
these channels is low, the observed CO morphology is difficult to reproduce in
detail with a simple, azimuthally symmetric disk model.  The CO emission for
this best-fit model is optically thin in both the J=2-1 and J=3-2 transitions,
even for the edge-on disk orientation, and therefore traces the full column 
density of disk material.

\begin{figure}
\centering
\includegraphics[scale=0.9]{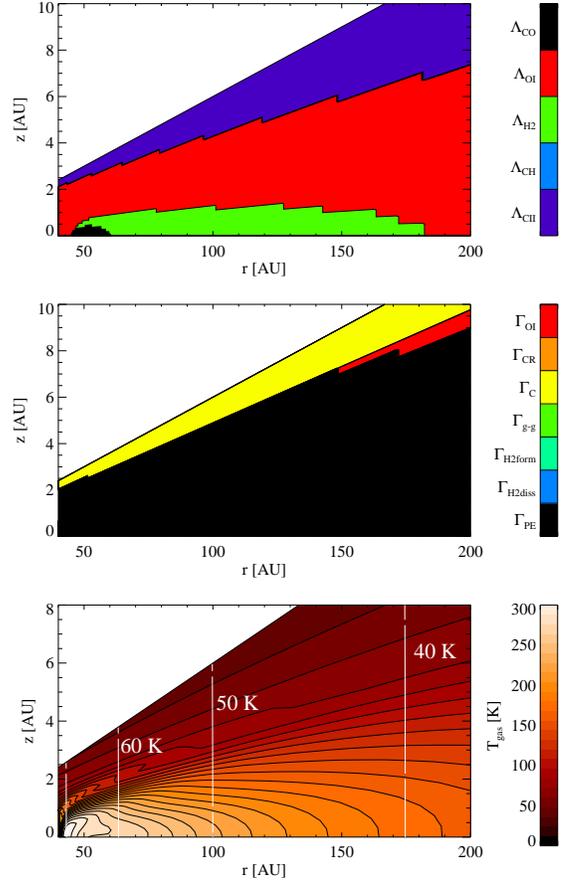}
\caption{
Two-dimensional gas temperatures in the best fit disk model ($M_{\rm disk}
=13$~M$_\earth$,
$\epsilon= 2.5$, $h=0.02$. Shown are the most important heating (top panel) 
and cooling (middle panel) processes as well as the gas temperature (bottom 
panel). The dust temperature, which depends only on radius, is overlaid in
white 
contour lines (steps of 10~K).
}
\label{gastemp}
\end{figure}

The densities in the disk are too low for efficient gas-dust coupling and 
thus the gas finds its own equilibrium temperature determined mainly by 
photoelectric heating and line cooling. The most important cooling lines from 
surface to midplane are [C\,{\sc ii}], [O\,{\sc i}], and H$_2$. CO abundances 
are only high in a region between 45 and 70~AU (Fig.~\ref{COoverview1}). 
Outside that region, CO cooling is less important for the energy balance. 
Fig.~\ref{gastemp} summarizes the most important heating and cooling processes 
and also shows the resulting gas temperature structure. The disk surface stays 
well below 100~K due to efficient fine structure line cooling. The molecular 
cooling is however less efficient in competing with the photoelectric heating 
from the large silicate grains \citep{kam01}, leading to temperatures of a few 
100~K in the disk interior.

\begin{figure}
\centering
\includegraphics[scale=0.7]{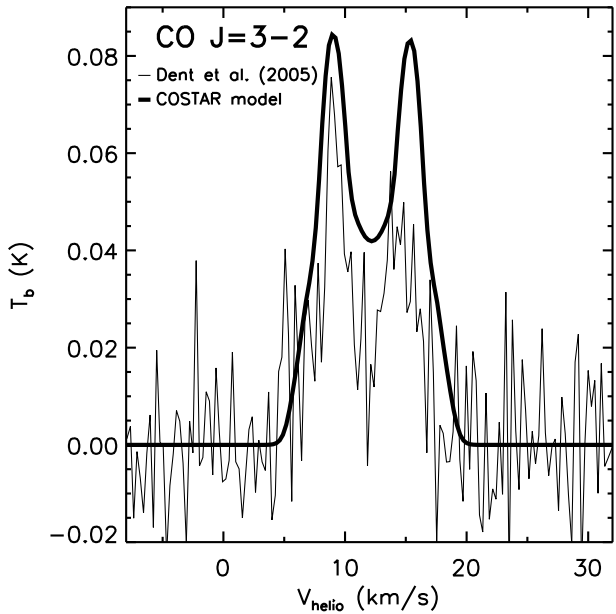}
\caption{
CO J=3-2 spectrum predicted for the model that provides the best fit to the 
resolved J=2-1 emission (heavy solid line), compared with the \citet{den05} 
JCMT CO J=3-2 spectrum (light solid line).  The x-axis shows heliocentric
velocity while the y-axis gives the JCMT main beam brightness temperature.  
}
\label{fig:spec}
\end{figure}

In order to test the robustness of the best-fit model to the gas properties,
we used this model to predict the CO J=3-2 spectrum.  It compares
favorably with the spectrum observed by \citet{den05}, reproduced in  
Figure~\ref{fig:spec}.  The heavy solid line shows the J=3-2 spectrum predicted 
from the best-fit disk model, while the light solid line shows the 
observed JCMT spectrum.  
Although the observed spectrum is noisy and likely 
subject to an absolute calibration uncertainty, the overall agreement is 
within $\sim 30$\%, which is very good given that the CO J=3-2 spectrum was 
not used {\em a priori} to determine these model parameters.

\section{Discussion}
\label{sec:dis}

The processes determining the amount and distribution of gas and dust in
transition disks like that around 49 Ceti are the same processes that shape 
the features of emergent planetary systems around these young stars. Resolved 
observations of individual disks in this phase are desirable to address such 
basic questions as when in the lifetime of a star its disk disperses, whether 
the gas clears before the dust, and whether the disk clears from the center or 
in a radially invariant manner.  

In the 49 Ceti system, the infrared dust properties appear similar to those of
a debris disk \citep{wah07}.  Yet observations presented here indicate that 
a substantial quantity of molecular gas persists in the outer disk, between
radii of 40 and 200 AU, where photochemistry from stellar and interstellar 
radiation dominates.  The lack of molecular gas emission interior to this 
radius as indicated by our observations, combined with the lack of dust emission
within a radius of 30 AU inferred by \citet{wah07}, implies that the 49 Ceti
system appears to be clearing its gas and dust from the center out.  
The mechanism responsible for this central clearing is not 
indicated; in general, the best-developed theories to explain this transitional
morphology are (1) central clearing through the influence of a massive planet 
and (2) photoevaporation by radiation from the central star.  

The clearing of gaps and inner holes has long been predicted as a consequence
of the formation of massive planets in circumstellar disks 
\citep[e.g.][]{lin86,bry99}.  In the case of 49 Ceti, the formation of a 
Jupiter-mass planet would be required at a distance of $\sim 40$ AU from the 
star, roughly the inner radius of the observed hole in the gas distribution.  
Such a scenario could also help to explain the size segregation of dust grains 
observed by \citet{wah07}; a predicted consequence of inner disk clearing 
by gravitational influence of a massive planet is a filtration of dust grains
by size, with only those below a certain threshold (typically 1-10 $\mu$m) 
accreted across the gap along with a reduced amount of gas \citep{ric06}.  
However, this scenario ultimately requires the accretion of substantial 
amounts of gas into the inner disk, and searches for molecular gas in the 
inner disk of 49 Ceti \citep{che06,car07} have not detected such a population. 
Another indication that an inner hole is likely caused by a massive planet
in formation would be non-axisymmetric features resulting from its
gravitational influence, such as spiral waves. While the CO emission from
49 Ceti does not appear asymmetric within the limits of the SMA observations,
more sensitive spatially resolved observations could address this hypothesis.

The absence of gas in the inner disk is, however, consistent with a 
photoevaporation scenario: as the photoevaporative wind produced by stellar
radiation becomes comparable to the accretion rate in the disk, material 
within the gravitational radius $R_g = GM_\star / c_s^2$ will quickly drain 
onto the star, leaving an evacuated inner hole free of gas and dust 
\citep[e.g.][]{hol94,ale06}.  The gravitational radius for 49 Ceti is roughly
20 AU, which is comparable to the inferred inner radius of 40 AU
for the outer disk.  The larger outer radius may in fact be consistent with
the later stages of photoevaporation, after the inner disk has become optically
thin to ultraviolet radiation and the inner disk radius slowly increases under 
the influence of the photoevaporative wind \citep{ale06}.  
\citet{ale07} propose a method of discriminating between inner holes caused
by photoevaporation and those caused by the formation of a giant planet, 
involving a simple comparison between two observables: the disk mass and the 
accretion rate.  As there is no measured accretion rate for 49 Ceti, we cannot
apply the criteria presented by these authors; however, we note that the low
disk mass does indeed fall within the parameter space consistent with 
a photoevaporative scenario.  Further observations are necessary to determine
the origin of the inner hole; in particular, stringent limits on the accretion
rate could suggest a photoevaporative mechanism. 

There are few disks which appear to be in a similar evolutionary stage to that
of 49 Ceti; a rare example is the disk around the A star HD 141569.  Like 49 
Ceti, it hosts a disk composed primarily of sub$\mu$m-size grains with infrared 
properties approaching those of a debris disk \citep{wah07,mar02}, while still 
retaining a substantial quantity of molecular gas with central region clear of 
gas emission, in this case out to a radius of $\sim11$ AU \citep{got06,bri07}.  
It exhibits a transitional SED \citep{mer04}, and observations of the 
rovibrational CO spectrum reveal gas with disparate rotational and vibrational 
temperatures \citep[200 K and 5000 K respectively; ][]{bri07}, indicative of UV 
fluorescence on the outer edges of an inner disk region cleared of gas and 
dust.  An analysis of the chemistry and gas properties of the system similar 
to the one presented here for 49 Ceti was conducted by \citet{jon06}. While 
the presence and extent of the inner hole are clearly indicated, the physical 
origin of this clearing is less obvious.  The Br$\gamma$ profile is indicative 
of substantial accretion, and \citet{bri07} deem a photoevaporative clearing 
mechanism unlikely due to the large column density outside the cleared region 
and the lack of evidence for a photoevaporative wind in the FUV \citep{mar05}.  
However, \citet{mer04} place a much lower limit of $10^{-11}$ M$_\sun$/yr on the
accretion rate, based on the assumed gas:dust ratio of 100 and the low optical
depth of the inner disk, which would be much more consistent with a 
photoevaporation scenario.  \citet{got06} note that the rough coincidence of
the inner rim of the disk with the gravitational radius suggests that 
photoevaporation in concert with viscous accretion is a likely cause for the 
inner disk clearing. 

Whatever the origin of their morphology, the observed gas and dust properties
indicate that the disks surrounding both 49 Ceti and HD 141569 appear to be in 
a transitional state of evolution during which the dust properties are 
beginning to appear more like those of a debris disk, while the gas is in the 
process of being cleared from the disk from the center out.  

\section{Conclusions}
\label{sec:con}

The SMA CO J=2-1 observations presented here provide the first spatially 
resolved observations of molecular gas in the 49 Ceti system.  The data reveal 
a surprisingly extended and complex molecular gas distribution in rotation 
about the central star, viewed approximately edge on and clear of molecular gas 
emission in the central region of the disk.  Modeling the disk structure and 
chemistry in this system indicates that the inner disk is entirely devoid of 
molecular gas due to irradiation by the central star, while a ring of molecular 
gas persists between 40 and 200 AU, subject to photodissociation 
at the inner edge by stellar radiation.  The disk model presented here 
reproduces well the observed properties of the system, including the resolved 
CO J=2-1 emission, the CO J=3-2 spectrum, and the spectral energy distribution.
With dust properties similar to those of a debris disk and a substantial 
reservoir of gas maintained in the outer disk, 49 Ceti appears to be a rare 
example of a system in a late stage of transition between a gas-rich 
protoplanetary disk and a tenuous, gas-free debris disk.

\acknowledgements
The authors would like to thank Bill Dent for providing the JCMT CO J=3-2
spectrum.  Partial support for this work was provided by NASA Origins
of Solar Systems Program Grant NAG5-11777.  A. M. H. acknowledges support
from a National Science Foundation Graduate Research Fellowship.

\end{document}